\documentclass[11pt]{article}

\begin{document}
\renewcommand{\baselinestretch}{1.3}

\title{Scale Symmetry Breaking From Total Derivative Densities 
and the Cosmological Constant Problem}

\author{Eduardo I. Guendelman \\
\small\it Department of Physics, Ben-Gurion University, Beer-Sheva, Israel  \\[-1.mm]
\small\it email: guendel@bgu.ac.il \\
${}$ \\
Hitoshi Nishino  and Subhash Rajpoot\\
\small\it California State University at Long Beach,\\[-1.mm]
\small\it Long Beach, California, USA  \\[-1.mm]
\small\it email: hnishino@csulb.edu, Subhash.Rajpoot@csulb.edu}

\maketitle

\begin{abstract}
The use in the action integral of  totally divergent densities 
in generally coordinate invariant theories can lead to interesting mechanisms
of spontaneous symmetry breaking of scale invariance. With dependence in the action 
on a metric independent density $\Phi$, in $4D$ , we can define 
$\Phi =\varepsilon^{\mu\nu\alpha\beta}\partial_{\mu}A_{\nu\alpha\beta}$ that gives a new interesting mechanism for breaking scale symmetry in  4-D theories of gravity plus matter fields, through the $A_{\nu\alpha\beta}$ equations of motion which lead to an integration constant the breaks the scale symmetry, while introducing terms of the form $eG ln K$ , $e$ being the determinant of the vierbein,
$G$ being the Gauss Bonnet scalar and $K$ being scalar functions of the fields transforming like 
$K \rightarrow cK $ (where c is a constant) under a scale transformation. Such a term is invariant only up to a total divergence and therefore leads to breaking of scale invariance due to gravitational instantons.
The topological density constructed out of gauge field strengths 
$\varepsilon^{\mu\nu\alpha\beta}F^a_{\mu\nu}F^a_{\alpha\beta}$ can be coupled to the dilaton field linearly to produce a scale invariant term up to a total divergence. The scale symmetry can be broken by Yang Mills instantons which lead to a very small vacuum energy for our Universe.
\end{abstract}

\section{Introduction}

The Cosmological Constant Problem (CCP) has evolved from the "Old Cosmological Constant Problem" \cite{W1},  where  physicist were concerned with explaining why the observed vacuum energy density of the universe is exactly zero, to different type of CCP since the evidence for the accelerating universe became evident, \cite{AU}.  We have therefore since the discovery of the accelerated universe a  "New Cosmological Constant Problem" \cite{W2},  the problem is now not to explain zero, but to explain a very small vacuum energy density.  

This new situation posed by the discovery of a very small vacuum energy density of the universe means that getting a zero vacuum energy density for the present universe is definitely not the full solution of the problem, although it may be a step towards its solution. In this respect one can take two inequivalent points of view:  I. The true vacuum of the theory has still have zero vacuum energy density, but we have not reached that point, so that is why we see a small vacuum energy density now. II. The true vacuum state of the theory has a non zero vacuum energy density and although there is a basic mechanism to drive the vacuum energy density to zero, some "residual" interaction is responsible for slightly shifting the vacuum energy density towards a small but non zero value. Here we are going to take this second point of view, that is, together with identifying a certain mechanism that drives the vacuum energy density to zero, we then consider a "residual" interaction that provides a small vacuum energy density. 

Interestingly enough both the basic mechanism that is able to drive the vacuum energy density of the universe to zero, for example the Two Measures Theory \cite{G1} and the small "residual" interaction that provides a small vacuum energy density (which will be activated through instantons) that we will explore here will involve totally divergent densities. 

First, concerning the basic mechanism to drive the vacuum energy density to zero,
we have studied models of the new class of
theories\cite{G1} and  based on the idea that the action
integral may contain the new metric-independent measure of
integration. For example, in four dimensions the new measure can
be built from a three index field as in
 $\Phi =\varepsilon^{\mu\nu\alpha\beta}\partial_{\mu}A_{\nu\alpha\beta}$ 
or of four scalar fields $\varphi_a$, ($a=1,2,3,4$)
\begin{equation}
\Phi
=\varepsilon^{\mu\nu\alpha\beta}\varepsilon_{abcd}\partial_{\mu}\varphi_a
\partial_{\nu}\varphi_b\partial_{\alpha}\varphi_c
\partial_{\beta}\varphi_d.
\label{Phi}
\end{equation}

These two representations give the same results.
There is another inequivalent choice  for constructing an alternative measure using four Lorentz vectors and preserving  local Lorentz invariance\cite{LLI}, this third approach opens new possibilities, not fully explored in details. 

$\Phi$ is a scalar density under general coordinate
transformations and the action can be chosen in the form $S = \int
L\Phi d^{4}x$. This has been applied to three different
directions: I. Investigation of the four-dimensional gravity and matter fields
models containing the new measure of integration that appears to
be promising for resolution of the dark energy and dark matter
problems, the fifth force problem, etc. II. Studying new type of string and brane models based on the use
of a modified world-sheet/world-volume integration measure \cite{G16}, \cite{G17}. It
allows new types of objects and effects like for example:
spontaneously induced string tension; classical mechanism for a
charge confinement; Weyl-conformally invariant light-like (WILL) brane \cite{KaluzaKlein} obtaining 
 promising results for black hole physics.III. Studying higher dimensional realization of the idea of the modified measure in the context of the Kaluza-Klein \cite{KaluzaKlein} and brane \cite{braneworld}
scenarios with the aim to solve the cosmological constant problem. Finally a mechanism for supersymmetry breaking has been found using a modified measure formulation of supergravity \cite{supergravity}.

We apply the
action principle to the action of the more general form
\begin{equation}
    S = \int L_{1}\Phi d^{4}x +\int L_{2}\sqrt{-g}d^{4}x,
\label{S}
\end{equation}
 including two Lagrangians $ L_{1}$ and $L_{2}$ and two volume elements ($\Phi d^{4}x$ and
$\sqrt{-g}d^{4}x$ respectively). To provide parity conservation,
one can  choose for example one of $\varphi_a$'s to be
pseudoscalar. Constructing the field theory with the action
(\ref{S}), we make only the basic additional  assumption that
$L_{1}$ and $L_{2}$ are independent of the measure fields
$\varphi_{a}$. Then the action (\ref{S}) is invariant under volume
preserving diffeomorphisms. Besides, it is
invariant (up to an integral of a total divergence) under the
infinite dimensional group of shifts of the measure fields
$\varphi_a$: $\varphi_a\rightarrow\varphi_a+f_{a}(L_{1})$,
where $f_{a}(L_{1})$ are arbitrary differentiable functions of
the Lagrangian  $L_{1}$. 
 We can proceed in the first order formalism where all fields,
including metric $g_{\mu\nu}$ (or vierbeins ${e}_{a\mu}$),
connection coefficients (or spin-connection $\omega_{\mu}^{ab}$)
and the measure fields $\varphi^{i}$ are independent dynamical
variables. All the relations between them follow from equations of
motion. The  field theory based on the listed assumptions we call
"Two Measures Theory" (TMT).

It turns out that the measure fields $\varphi_{a}$ affect the
theory only via the ratio of the two measures
\begin{equation}
\zeta\equiv \Phi /\sqrt{-g} \label{zeta}
\end{equation}
a scalar field that is determined by a constraint in
the form of an algebraic equation which is a consistency
condition of the equations of motion.

TMT models naturally avoid the 5th force problem \cite{5th} and naturally provide ground states with zero vacuum energy density. 

One should also notice that a similar structure to the Two Measures Theories has been found in the Hodge Dual formulation of Supergravity Theories \cite{G8}. 
The two measure theories have many points of similarity with ``Lagrange Multiplier Gravity (LMG)'' \cite{Lim2010,Capozziello2010}. 
In LMG there is a Lagrange multiplier field which enforces the condition that a certain function is zero. In the 
two measure theory this is equivalent to the constraint which requires some lagrangian to be constant. The two measure model presented here, as opposed to the LMG models of \cite{Lim2010,Capozziello2010} provide us with an arbitrary constant of integration. The introduction of constraints can cause Dirac fields to contribute to dark energy \cite{FDM} or scalar fields to behave like dust like in \cite{Lim2010} and this dust behaviour can be caused by the stabilization of a tachyonic field due to the constraint, accompanied by a floating dark energy component \cite{GSY, AG}.

Two measure theories can also be used to construct non singular "emergent" scenarios\cite{EMER} for the early  universe, that existed since arbitrarily large early times in the form of a stable Einstein Universe. This phase then gets transformed into an inflationary phase and subsequently into a slowly accelerated one. The requirement that the early phase exist can impose restrictions on the possible values of the cosmological constant at the end\cite{EMERCC}.

\section{Total Derivative Densities, 2nd order Form and S.S.B of Scale Symmetry from Instantons}
 
Coupling to other densities which are total derivatives provide us with other possibilities to break the scale symmetry due to non trivial boundary conditions provided by instanton solutions.

For example introducing $eGln K$,  where e is the determinant of the vierbein, $G$ is the Gauss Bonnet scalar and $K$ is a function of the fields that  transforms as $K \rightarrow cK$ under global scale invariance, $c$ being a constant, we obtain invariance up to a total divergence, since $e G$, is a total divergence, $G$ being given by,

\begin{equation}
 G= R^{\mu\nu\alpha\beta}R_{\mu\nu\alpha\beta}- 4R^{\mu\nu}R_{\mu\nu} + R^{2}
\label{GaussBonnet}
\end{equation}

In \cite{SchmiSing} terms of the type $eGln G$ were considered.
Here we  assume the second order formulation, which means the connection is assumed to be the Christoffel symbol, a well known function of the metric, unlike the first order formulation, which we will use in the next sections, where this assumption will be removed (and where the connection will be considered as a dynamical variable independent a priori from the metric). In this case, we achieve invariance under scale invariance transformations up to a total derivative. This total derivative can give a non trivial contribution under the presence of gravitational instantons \cite{GravIns}.

Likewise, if gauge fields are considered, a coupling of the dilaton to $\varepsilon^{\mu\nu\alpha\beta}F^a_{\mu\nu}F^a_{\alpha\beta}$ (here $F^a_{\alpha\beta}$ represents a non abelian field strength) also produces invariance up to a total derivative and if Yang Mills instantons \cite{YMIns} are considered, this could lead to physical consequences, since these total derivatives can contribute in the functional integral.

Using $\zeta$ as defined before and choosing a scaling assignment for the measure fields and for the metric that keeps $\zeta$ invariant, we define a general action which is scale invariant and that allows non linear $\zeta$ dependence and which furthermore assumes the second order formulation
\begin{equation}
 S = \int L d^{4}x
\label{nonlinearaction1}
\end{equation}
where
\begin{eqnarray}
 L = -\frac{e}{2} F_1 (\zeta) \sigma^{2}R(e) -e \sigma^{4} F_2 (\zeta)  +  e F_3 (\zeta)\frac{1}{2}g^{\mu\nu}\sigma_{,\mu}\sigma_{,\nu} + \alpha e R ^{2} +
 \beta e R^{\mu\nu}R_{\mu\nu} \nonumber  \\\
  -\frac{e}{4}F^a_{\mu\nu}F^{a\mu\nu} +e\Sigma_i \gamma_i G ln( W_i)
 +\delta ln(\sigma) \varepsilon^{\mu\nu\alpha\beta}F^a_{\mu\nu}F^a_{\alpha\beta}   
\end{eqnarray}

 The $W_i$ are functions of fields that under the scale transformation 
 $e^{m}_\mu \rightarrow e^\Lambda e^{m}_\mu$ undergo a transformation $W_i\rightarrow e^{k_i\Lambda}W_i$
 where $k_i$ characterizes any particular $W_i$. This action is then invariant up to a total derivative under the scale transformations $e^{m}_\mu \rightarrow e^\Lambda e^{m}_\mu$,  
$\varphi_a \rightarrow e^\Lambda \varphi_a$ and
$\sigma \rightarrow e^{-\Lambda} \sigma$.

 The equations of motion that follow from the variation of the $\varphi_a$ fields are
 \begin{eqnarray}
 A^{\beta}_d \partial_\beta (-\frac{1}{2} F_1 ^{\prime}(\zeta) \sigma^{2}R(e) - \sigma^{4} F_2^{\prime} (\zeta)  +   F_3^{\prime}(\zeta)\frac{1}{2}g^{\mu\nu}\sigma_{,\mu}\sigma_{,\nu}) = 0  \\\
A^{\beta}_d =  \varepsilon^{\mu\nu\alpha\beta}\varepsilon_{abcd}\partial_{\mu}\varphi_a
\partial_{\nu}\varphi_b\partial_{\alpha}\varphi_c
   \end{eqnarray}

 Since $det(A^{\beta}_d)$ is proportional to $\Phi^3$, we obtain that if $\Phi \neq 0$
 \begin{eqnarray}
 -\frac{1}{2} F_1 ^{\prime}(\zeta) \sigma^{2}R(e) - \sigma^{4} F_2^{\prime} (\zeta)  +   F_3^{\prime}(\zeta)\frac{1}{2}g^{\mu\nu}\sigma_{,\mu}\sigma_{,\nu} = C_0= constant
\end{eqnarray}

Since the right hand side is a constant and the left hand side transforms under scale transformations, we obtain spontaneous breaking of scale symmetry therefore.

Separately from this, the terms $e\Sigma_i \gamma_i G ln( W_i)$ and
 $\delta ln(\sigma) \varepsilon^{\mu\nu\alpha\beta}F^a_{\mu\nu}F^a_{\alpha\beta}$ which are invariant under a scale transformation up to a total derivative, can produce a breaking of global scale invariance once instanton contributions (both gravitational and Yang Mills) are considered.

\section{A simple TMT model allowing for a small vacuum energy density due to Instantons}
    We will study now the dynamics of a scalar field $\phi$ interacting
with gravity as given by the following action, 

\begin{equation}\label{e9}
S =    \int L_{2} \sqrt{-g}   d^{4} x +  \int L_{1} \Phi d^{4} x + \int N \phi \varepsilon^{\mu\nu\alpha\beta}F^a_{\mu\nu}F^a_{\alpha\beta} d^{4} x
\end{equation}
\begin{equation}\label{e10}
L_{2} = U(\phi)-\frac{1}{4}F^a_{\mu\nu}F^{a\mu\nu}
\end{equation}

\begin{equation}\label{e11}
L_{1} = \frac{-1}{\kappa} R(\Gamma, g) + \frac{1}{2} g^{\mu\nu}
\partial_{\mu} \phi \partial_{\nu} \phi - V(\phi)
\end{equation}
\begin{equation}\label{e12}
R(\Gamma,g) =  g^{\mu\nu}  R_{\mu\nu} (\Gamma) , R_{\mu\nu}
(\Gamma) = R^{\lambda}_{\mu\nu\lambda}
\end{equation}
\begin{equation}\label{e13}
R^{\lambda}_{\mu\nu\sigma} (\Gamma) = \Gamma^{\lambda}_
{\mu\nu,\sigma} - \Gamma^{\lambda}_{\mu\sigma,\nu} +
\Gamma^{\lambda}_{\alpha\sigma}  \Gamma^{\alpha}_{\mu\nu} -
\Gamma^{\lambda}_{\alpha\nu} \Gamma^{\alpha}_{\mu\sigma}.
\end{equation}

Notice that because of the $N$ coupling, such action does violate parity, since the field $\phi$ has associated potentials which are not even under $\phi \rightarrow -\phi$, so, assigning to this field a pseudo-scalar nature does not make the model parity or $CP$ conserving. We take $\phi$ as a scalar and the parity and $CP$ violating $N$
term to be very small. We notice at this point that this is consistent with 't Hooft naturalness condition. Indeed 't Hooft \cite{THOOFT}  states that "at any energy scale $\mu$, a physical parameter 
$\alpha_i(\mu)$ is allowed to be very small only if the replacement $\alpha_i(\mu)=0$ would increase the symmetry of the system". This is indeed exactly what happens here, for $N=0$ parity and $CP$ symmetries are restored in the scalar sector, so that allows us to say that the requirement of a small $N$, which is indeed needed if we want a small resulting vacuum energy density, is justified.

    In the variational principle $\Gamma^{\lambda}_{\mu\nu},
g_{\mu\nu}$, the measure fields scalars
$\varphi_{a}$ and the "matter" - scalar field $\phi$ are all to be treated
as independent
variables although the variational principle may result in equations that
allow us to solve some of these variables in terms of others. Treating the connection as independent a priori as independent of the metric is what is referred to as "first order formalism", as opposed to assuming that the connection is given by the Christoffel symbol or "second order formulation". It should be pointed out that the characterization of these two procedures as merely different formalisms is not correct, indeed, except for the special (although very important case) of General Relativity, these two procedures originate inequivalent theories.

 We can  have  global
scale invariance in this model for very special exponential form for the $U$ and $V$ potentials. Indeed, if we perform the global
scale transformation ($\theta$ =
constant)
$g_{\mu\nu}  \rightarrow   e^{\theta}  g_{\mu\nu}$,
then there is invariance provided  $V(\phi)$ and $U(\phi)$ are of the
form  \cite{G2} (where the case without coupling to the gauge fields topological density was studied, see also \cite{GKatz} for the generalization in the case square curvature terms are included)
\begin{equation}\label{e19}
V(\phi) = f_{1}  e^{\alpha\phi},  U(\phi) =  f_{2}
e^{2\alpha\phi}
\end{equation}
and $\varphi_{a}$ is transformed according to
$\varphi_{a}   \rightarrow   \lambda_{ab} \varphi_{b}$
which means
$\Phi \rightarrow det( \lambda_{ab}) \Phi \\ \equiv \lambda
\Phi $
such that
$\lambda = e^{\theta}$
and
$\phi \rightarrow \phi - \frac{\theta}{\alpha}$.

We will now work out the equations of motion after introducing $V(\phi)$ and $U(\phi)$
and see how the integration of the equations of motion allows the spontaneous breaking of the
scale invariance.  

    Let us begin by considering the equations which are obtained from
the variation of the fields that appear in the measure, i.e. the
$\varphi_{a}$
fields. We obtain,
$A^{\mu}_{a} \partial_{\mu} L_{1} = 0$
where  $A^{\mu}_{a} = \varepsilon^{\mu\nu\alpha\beta}
\varepsilon_{abcd} \partial_{\nu} \varphi_{b} \partial_{\alpha}
\varphi_{c} \partial_{\beta} \varphi_{d}$. As in the previous section, if $\Phi\neq 0$.
we obtain that $\partial_{\mu} L_{1} = 0$,
 or that
\begin{equation}\label{e25}
L_{1} = \frac{-1}{\kappa} R(\Gamma,g) + \frac{1}{2} g^{\mu\nu}
\partial_{\mu} \phi \partial_{\nu} \phi - V = M
\end{equation}
where M is constant. Notice that this equation breaks spontaneously the global scale invariance of the theory, 
since the left hand side has a non trivial transformation under the scale transformations, while the right
hand side is equal to $M$, a constant that after we integrate the equations is fixed, cannot be changed and therefore
for any $M\neq 0$ we have obtained indeed, spontaneous breaking of scale invariance.

    Considering now the variation with respect to $g^{\mu\nu}$, we
obtain
\begin{equation}\label{e31}
\Phi (\frac{-1}{\kappa} R_{\mu\nu} (\Gamma) + \frac{1}{2} \phi,_{\mu}
\phi,_{\nu}) - \frac{1}{2} \sqrt{-g} U(\phi) g_{\mu\nu} + \sqrt{-g} (F^a_{\mu\alpha}F^{a\alpha}_\nu -
\frac{1}{4}g_{\mu\nu}F^a_{\alpha \beta}F^{a \alpha \beta}) = 0
\end{equation}
solving for $R = g^{\mu\nu} R_{\mu\nu} (\Gamma)$  from eq.\ref{e31}  and introducing
in eq.\ref{e25}, we obtain
a constraint that allows us to solve for the ratio of the two measures $\zeta$,
\begin{equation}\label{e33}
\zeta = \frac{2U(\phi)}{M+V(\phi)}.
\end{equation}

    To get the physical content of the theory, it is best to
consider variables that have a well defined dynamical interpretation. The original
metric does not has a non zero canonical momenta in the first order formalism as no derivatives
of such metric appear in the lagrangian, all derivatives appear in the connections, which are the fundamental dynamical
variables of the theory. The
canonical momenta of those connections are functions of $\overline{g}_{\mu\nu}$, given by,

\begin{equation}\label{e34}
\overline{g}_{\mu\nu} = \zeta g_{\mu\nu}
\end{equation}

and $\zeta$  given by eq.\ref{e33}. Interestingly enough, working with $\overline{g}_{\mu\nu}$
is the same as going to the "Einstein Conformal Frame".
Defining $\Sigma^{\lambda}_{\mu\nu} =
\Gamma^{\lambda}_{\mu\nu} -\{^{\lambda}_{\mu\nu}\}$
where $\{^{\lambda}_{\mu\nu}\}$   is the Christoffel symbol, it turns out that in terms of $\overline{g}_{\mu\nu}$   the non
Riemannian contribution $\Sigma^{\alpha}_{\mu\nu}$
disappears from the equations. This is because the connection
can be written as the Christoffel symbol of the metric
$\overline{g}_{\mu\nu}$ .
In terms of $\overline{g}_{\mu\nu}$ the equations
of motion for the metric can be written then in the Einstein
form (we define $\overline{R}_{\mu\nu} (\overline{g}_{\alpha\beta}) =$
 usual Ricci tensor in terms of the bar metric $= R_{\mu\nu}$ and
 $\overline{R}  = \overline{g}^{\mu \nu}  \overline{R}_{\mu\nu}$ )
\begin{equation}\label{e35}
\overline{R}_{\mu\nu} (\overline{g}_{\alpha\beta}) - \frac{1}{2}
\overline{g}_{\mu\nu}
\overline{R}(\overline{g}_{\alpha\beta}) = \frac{\kappa}{2} T^{eff}_{\mu\nu}
(\phi)
\end{equation}

\begin{equation}\label{e36}
T^{eff}_{\mu\nu} (\phi) = \phi_{,\mu} \phi_{,\nu} - \frac{1}{2} \overline
{g}_{\mu\nu} \phi_{,\alpha} \phi_{,\beta} \overline{g}^{\alpha\beta}
+ \overline{g}_{\mu\nu} V_{eff} (\phi) + F^a_{\mu\alpha}F^{a\alpha}_\nu -
\frac{1}{4}g_{\mu\nu}F^a_{\alpha \beta}F^{a \alpha \beta}
\end{equation}

\begin{equation}\label{e37}
V_{eff} (\phi) = \frac{1}{4U(\phi)}  (V+M)^{2}.
\end{equation}

    Using the metric $\overline{g}^{\alpha\beta}$ the equation of the
field $\phi$ becomes

$\frac{1}{\sqrt{-\overline{g}}} \partial_{\mu} (\overline{g}^{\mu\nu}
\sqrt{-\overline{g}} \partial_{\nu}
\phi) + V^{\prime}_{eff} (\phi) + N \frac{\varepsilon^{\mu\nu\alpha\beta}F^a_{\mu\nu}F^a_{\alpha\beta}}{\sqrt{-\overline{g}}}  = 0$. In the case $N=0$, the vacuum is obtained for  $V + M = 0$, where $V_{eff}  = 0$ and $V^{\prime}_{eff}
= 0$ also, provided $V^{\prime}$ is finite and $U \neq 0$ there.
This means the vacuum with zero cosmological constant
state
is achieved without any sort of fine tuning. That is, independently
of whether we add to $V$ a constant piece, or whether we change
the value of $M$, as long as there is still a point
where $V+M =0$, then still $ V_{eff}  = 0$ and $V^{\prime}_{eff} = 0$.
This is the basic feature
that characterizes the TMT and allows it to solve the 'old'
cosmological constant problem.

The consideration of $N \neq 0$ changes this picture, because of the additional $N$ term, the scalar 
field may not sit exactly at the minimum of the effective potential. The  $N$ term acts indeed as an external source driving the scalar field away from such point and we expect indeed this to be the case when considering the effect of instantons.

    If $V(\phi) = f_{1} e^{\alpha\phi}$  and  $U(\phi) = f_{2}
e^{2\alpha\phi}$ as
required by scale
invariance, we obtain from the expression in eq.(\ref{e37})
\begin{equation}\label{e40}
    V_{eff}  = \frac{1}{4f_{2}}  (f_{1}  +  M e^{-\alpha\phi})^{2}
\end{equation}

    Since we can always perform the transformation $\phi \rightarrow
- \phi$ we can
choose by convention $\alpha > 0$. We then see that as $\phi \rightarrow
\infty, V_{eff} \rightarrow \frac{f_{1}^{2}}{4f_{2}} =$ const.
providing an infinite flat region. Also a minimum is achieved at zero
cosmological constant for the case $\frac{f_{1}}{M} < 0$ at the point
\begin{equation}\label{phimin1}
\phi_{min}  =  \frac{-1}{\alpha} ln \mid\frac{f_1}{M}\mid.
\end{equation}

We are now ready to consider the effect of the $N$ term, which drives the scalar field away from the absolute minimum of the effective potential and therefore from the zero vacuum energy density vacuum.
The $N$ term does not contribute to the energy momentum tensor, since its contribution to the action is metric independent, but it does affect the vacuum energy density, because it can push the dilaton away from the minimum of $V_{eff}$.

Then the vacuum state is found now when
\begin{equation}\label{e38}
 V^{\prime}_{eff} (<\phi>) + N <\frac{\varepsilon^{\mu\nu\alpha\beta}F^a_{\mu\nu}F^a_{\alpha\beta}}{\sqrt{-\overline{g}}}>  = 0.
\end{equation}

To proceed with the estimation of the vacuum energy density, it is necessary to handle the expectation value of the topological density. We may proceed by analogy with the axion field \cite{axion} mass generation calculation. In the case of the axion, the expectation value of the topological density of the QCD gauge fields is shown to lead to the generation of a mass. 

The expectation value depends on the expectation value of the dilaton and on the theta parameter of the QCD vacuum. Let us denote the value of the dilaton field for which this expectation value vanishes $\phi_0$ . There is also, as in the case of the axion a mass generation around that value, and parametrizing $N = \psi \alpha_s / 8 \pi  f_\phi$, ($\psi$ being of order one ), to make contact with the way the interaction of an axion to the topological density is presented \cite{Peccei}. Then this allows us to utilize the results known for axion mass generation from instantons to obtain in our case the dilaton  mass generated from instantons of the order of
$m_\phi =  0.6 \frac{10^7Gev}{ f_\phi}ev $. This mass concerns oscillations around some unknown value $\phi_0$.
So we have a resulting effective potential
\begin{equation}\label{instantoncor}
 V_{eff-TOTAL} (\phi) =  V_{eff} (\phi) + m^2_\phi (\phi - \phi_0)^2/2
\end{equation}

The effect of the new mass term due to instanton effects is assumed to be very small, so we solve the value of the scalar field by a perturbative approach, where the main effect is given by the minimum of the original potential (that is when $m_\phi  = 0$) plus a perturbation due to this new term, which we will denote $\delta \phi$
\begin{equation}\label{phicorrected}
 <\phi> = \frac{-1}{\alpha} ln \mid\frac{f_1}{M}\mid + \delta \phi
\end{equation}
 
considering that for the value $\frac{-1}{\alpha} ln \mid\frac{f_1}{M}\mid$, $V_{eff}=0$ and that the same value represents the minimum of 
$V_{eff} (\phi)$ and making use of the fact that at that point $V''_{eff}= f^2_1 \alpha^2/2f_2 $.
So  keeping  only quadratic terms in $\delta \phi$ we obtain
\begin{equation}\label{instantoncor1}
 V_{eff-TOTAL} (\delta \phi) =  f^2_1 \alpha^2 (\delta \phi)^2/4f_2 + 
 m^2_\phi (\delta \phi - \phi'_0)^2 /2
\end{equation}
where $ \phi'_0 =  \phi_0 -  \frac{1}{\alpha} ln \mid\frac{f_1}{M}\mid  $. Then
$ V_{eff-TOTAL} (\delta \phi)$ is   minimized
for

\begin{equation}\label{instantoncor1}
\delta \phi =  m^2_\phi \phi'_0 / ( m^2_\phi + f^2_1 \alpha^2 /2f_2   ) 
\end{equation}

We are now interested in calculating the non vanishing vacuum energy density that is produced by this  
$\delta \phi$.  For this purpose it is very important to notice that only $ V_{eff}$ enters in the energy momentum tensor, the "force" caused by the $K$ term does not enter in the energy momentum tensor, its gravitational effect is indirect, by shifting the position of  the vacumm, but the vacuum energy density is still just $ V_{eff}$, which for the shifted value has now the non vanishing value

\begin{equation}\label{instantoncor2}
  \Lambda_{eff} = V_{eff} (\delta \phi) = m_\phi^4 (\phi'_0)^2 \alpha^2 f^2_1 / 4 f_2 ( m^2_\phi + f^2_1 \alpha^2 /2f_2  )^2
\end{equation}
It is very important to point out that  $m_\phi$ does not represent the mass of the dilaton, the mass of the dilaton is non zero even when the $m_\phi$ contribution 
is not considered. The fact that $m_\phi$ is small does not mean at all that the mass of the dilaton is small, it only means that the force that shifts the vacuum from the zero cosmological constant state is very small, but the mass of the dilaton at that vacuum can be large (may be as large as the Higgs mass).

\section{Discussion and Conclusions} The consideration of densities which are total derivatives allows to 
obtain new effects not easily available otherwise. It allows the breaking of symmetries, as it is known in the case of chiral symmetry, the anomaly allows its breaking via instantons. Similar effect can be exploited for the case of scale symmetry, where we have considered both the effects of a modified measure, which is metric independent and a total derivative, the Gauss Bonnet scalar and of the topological Yang Mills density.
With respect to the cosmological constant problem, our main result is that it is best to consider this problem in the context of the Two Measure Theory. 

This theory corresponds to the case when the only density which is a total derivative that is considered is a measure that is independent of the metric
 and it appears in a linear form in the action. This feature is protected by an infinite dimensional symmetry. In this case, the effective potential in the Einstein frame comes out as a perfect square so that one can naturally obtain the zero of the effective potential at the same point as the potential vanishes. This in fact is independent of the choice of scale invariance for our theory, but for the scale invariant choice, we obtain indeed SSB of this scale invariance.
 
In any case, the choice of a perfect square for the effective potential is a result that we take as an 
answer to the "old cosmological constant problem" or why we get zero for the vacuum energy density 
in the ground state of the theory. It is interesting to note that the perfect square structure for the effective potential for a multi field case (of Higgs and dilaton)was assumed in the the paper 
\cite{cosmon} which postulated the "cosmon" (we would call it dilaton) as a driver to zero cosmological constant. The TMT supports this assumption indeed.

The next thing we tackle is the "new cosmological constant problem". We find 
that the coupling of the dilaton to a Yang Mills topological density due to the effect of Yang Mills instantons can shift the value of the scalar field away from the minimum with zero vacuum energy of  the
simpler TMT (that does not couple to the Yang Mills topological density).

It is interesting to note that coupling of the dilaton to the Yang Mills topological density is 
absolutely metric independent and therefore such coupling does not enter in the energy momentum 
tensor. It has however the capability of shifting the dilaton away from the point of zero cosmological constant, that leads to a very small vacuum energy density in the vacuum.

The smallness of the vacuum energy density in the vacuum, depends on the smallness of $m_\phi$, which in turn, depends on the smallness of $N$. Notice that $m_\phi$ does not represent the mass of the dilaton, which can be many orders of magnitude bigger that $m_\phi$. This parater $m_\phi$ is related to the "force" that pushes the dilaton awy from the vacuum with zero vacuum energy density.

We notice that taking $N$ small is consistent with 't Hooft naturalness condition \cite{THOOFT}  which states that "at any energy scale $\mu$, a physical parameter 
$\alpha_i(\mu)$ is allowed to be very small only if the replacement $\alpha_i(\mu)=0$ would increase the symmetry of the system". This is indeed exactly what happens here, for $N=0$ parity and $CP$ symmetries are restored in the scalar sector, so that allows us to say that the requirement of a small $N$ is justified and this in turn is indeed exactly what is needed if we want a small resulting vacuum energy density. 

Finally some words should be said concerning the motivation for the Two Measures Theories. It appears that the way we have applied it in this paper, the Two Measure Theory represents the minimal extension of General Relativity that also allows us to handle the new cosmological constant problem.  The idea of two measures can be motivated in a number of different ways also, we refer for a more complete discussion to \cite{Foundations} for a list of different possible origins for these kind of models. These include space-time filling branes which naturally lead to a measure that corresponds to a jacobian of the mapping of two spaces, the brane models, where naturally two types of contributions appear, the brane and the bulk, using different measures of integrations, etc.

\textbf{Acknowledgements.} EG gratefully acknowledges the hospitality at UCLA and CSULB where this work was carried out. We want to thank Zvi Bern and Roberto Peccei for very useful conversations.

\end{document}